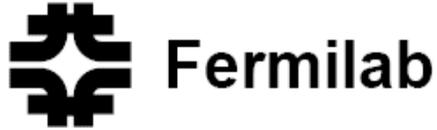

# OPTIMIZATION OF THE TARGET SUBSYSTEM FOR THE NEW G-2 EXPERIMENT[*][†]

C. Yoshikawa[#], C. Ankenbrandt
Muons, Inc., Batavia, IL 60510, U.S.A
A. Leveling, N.V. Mokhov, J. Morgan, D. Neuffer, S. Striganov
Fermilab, Batavia, IL 60510, U.S.A.

## Abstract

A precision measurement of the muon anomalous magnetic moment, $a_\mu = (g-2)/2$, was previously performed at BNL with a result of 2.2 - 2.7 standard deviations above the Standard Model (SM) theoretical calculations. The same experimental apparatus is being planned to run in the new Muon Campus at Fermilab, where the muon beam is expected to have less pion contamination and the extended dataset may provide a possible 7.5σ deviation from the SM, creating a sensitive and complementary benchmark for proposed SM extensions. We report here on a preliminary simulation study of the target subsystem where the apparatus is optimized for pions that have favourable phase space to create polarized daughter muons around the magic momentum of 3.094 GeV/c, which is needed by the downstream g 2 muon ring.

---

[*]Work supported by Fermi Research Alliance, LLC under contract No. DE-AC02-07CH11359 with the U.S. Department of Energy.
[†]Presented paper at International Particle Accelerator Conference 2012, May 20-25, 2012, New Orleans, U.S.A.
[#]cary.yoshikawa@muonsinc.com

# OPTIMIZATION OF THE TARGET SUBSYSTEM FOR THE NEW G-2 EXPERIMENT[*]

C. Yoshikawa[#], C. Ankenbrandt, Muons, Inc., Batavia, IL 60510, U.S.A.
A. Leveling, N.V. Mokhov, J. Morgan, D. Neuffer, S. Striganov, Fermilab, Batavia, IL 60510, U.S.A

*Abstract*

A precision measurement of the muon anomalous magnetic moment, $a_\mu = (g-2)/2$, was previously performed at BNL with a result of 2.2 - 2.7 standard deviations above the Standard Model (SM) theoretical calculations. The same experimental apparatus is being planned to run in the new Muon Campus at Fermilab, where the muon beam is expected to have less pion contamination and the extended dataset may provide a possible 7.5σ deviation from the SM, creating a sensitive and complementary benchmark for proposed SM extensions. We report here on a preliminary simulation study of the target subsystem where the apparatus is optimized for pions that have favourable phase space to create polarized daughter muons around the magic momentum of 3.094 GeV/c, which is needed by the downstream g 2 muon ring.

## INTRODUCTION

The New g-2 Experiment at Fermilab [1] aims to measure the muon anomalous magnetic moment to a precision of ±0.14 ppm ─ a fourfold improvement over the 0.54 ppm precision obtained in the g-2 BNL E821 experiment [2]. The present discrepancy, $\Delta a_\mu$(Expt. ─ SM) = $(255\pm80)\times10^{-11}$, is already suggestive of possible new physics contributions to the muon anomaly. Assuming that the current theory error of $49\times10^{-11}$ is reduced to $30\times10^{-11}$ on the time scale of the completion of our experiment, a future $\Delta a_\mu$ comparison would have a combined uncertainty of $\approx 34 \times 10^{-11}$, resulting in a 7.5σ deviation from the Standard Model, which will be a sensitive and complementary benchmark for proposed extensions to the Standard Model. Most of the improvement will be due to increased statistics and thus it is essential to maximize production of useful pions that create polarized muons which are in the acceptance of the g-2 muon storage ring. Furthermore, cost considerations favour a design that reuses the existing pbar production subsystem that worked well during the Tevatron operation. Hence, the pion production subsystem will begin with the pbar production subsystem scaled from 8 GeV (kinetic energy) protons to 3.1 GeV/c pions.

## THE LAYOUT

A graphical representation of the Fermilab pbar production target subsystem is shown in **Figure 1** as implemented in Ref. [3] in the MARS15 code [4]. The proton beam with kinetic energy of 8 GeV impinges on the default target, which is a vertical cylinder (in-out of top view in Figure **1**) composed primarily of inconel with a chord for the proton beam of ~7.5 cm. Pions produced in the target will be focused by the Li lens (yellow) that is 16 cm long, 1 cm in radius, and has a magnetic field gradient of 256.25 T/m, where the gradient has been scaled for 3.1 GeV/c pions to maintain proper focusing, while keeping the same focusing distance between centers of the target and Li lens of 25.16 cm. The focused pion beam is then collimated and bent through a pulsed magnet (PMAG) with a dipole field of 0.542219 T, also scaled for the 3.1 GeV/c pion beam, and bends the reference by 3 degrees to provide momentum selection.

A transition in our simulation between MARS that provides reliable particle generation and G4beamline [5] that is used for particle tracking, pion decay into muons, and effect of beam particles interacting with the beam line elements is shown in Figure 2. The MARS particle tracks that hit the virtual detector are converted and propagated in G4beamline through a set of four quads that refocuses the beam after the three degree bend from the PMAG. Figure 3 shows 100 such particle tracks traversing the four quads.

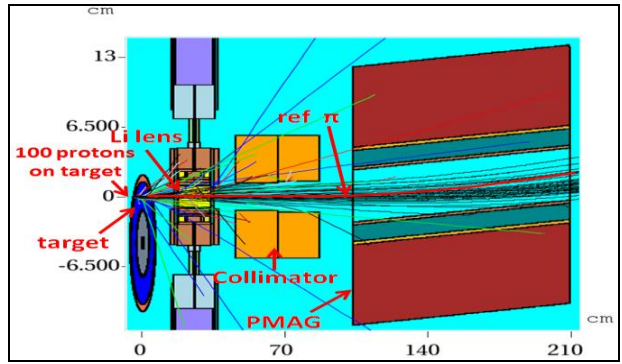

Figure 1: Zoomed in top view of pbar target subsystem.

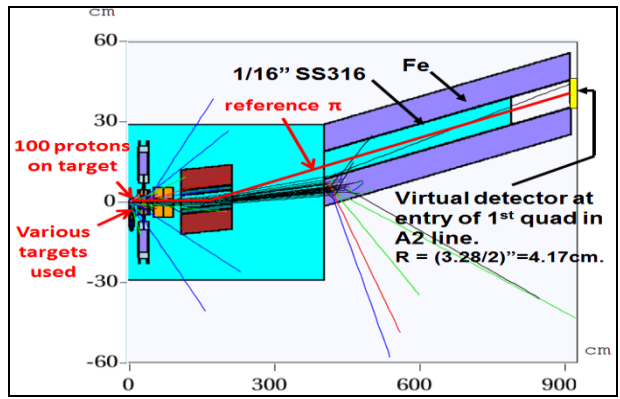

Figure 2: Top view of pbar target subsystem.

___________________
*Work supported by Fermi Research Alliance, LLC, under contract No. DE-AC02-07CH11359 with the U.S. Department of Energy
[#]cary.yoshikawa@muonsinc.com

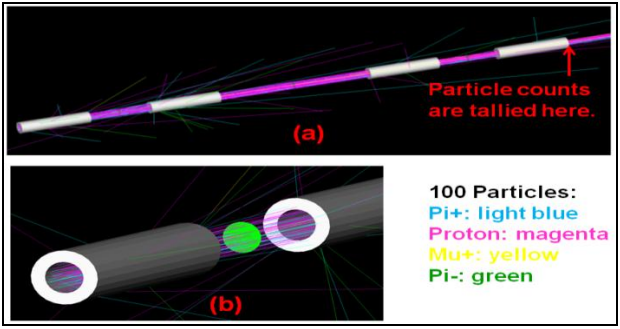

Figure 3: Particles after the conversion into G4beamline and propagated through the 4 quads. (a) Particle yields are tallied at end of the fourth quad with acceptance cuts appropriate for downstream elements. (b) Zoomed in view of particle trajectories between quads that are seen by a virtual detector (green).

The particle yields destined for the g-2 muon ring are estimated by particles simulated to the end of the fourth quad in the Fermilab A2 line as shown in Figure 3(a) and applying the acceptance of those downstream elements, which are:

- P(pi+) = 3.15588 GeV/c±2% ($1.02P_{magic}\pm0.02P_{magic}$)
- 40π mm-mrad in each transverse dimension

## THE OPTIMIZATION

The parameters investigated in this optimization study are the incident proton beam spot size, the length of the target, and the orientation of the target. We considered two spot sizes for the proton. One is what we expect from a simple scaling from 120 GeV operation to 8 GeV. The other is the smallest we believe that can be achieved. Spot size information on both is provided in Table 1. Prior to the start of this study, we benefited from an earlier investigation [6] that showed a smaller beam spot on long thin cylindrical targets improved yield of useful pions over the default proton beam spot size on the default pbar target. From that study, Figure 4 shows that in the range for the β-function at the target, the yield appears to improve with longer targets, while

Figure 5 illustrates a weak dependence on the target thickness. The present investigation extends that study to consider a more practical target that accommodates cooling. The present design is a thin walled cylinder of inconel where the proton beam impinges on the thin wall in the direction parallel to the axis of the cylindrical target as shown in Figure 6(a & c). The cylindrical target was approximated with slab targets in vertical and horizontal orientations that correspond to where along the azimuth the beam hits the target, as identified in Figure 6. The dimensions of these slab targets in this study along with a reference solid cylindrical target simulated in the earlier study [6] and the default pbar target are given in Table 2. The maximum length under consideration has increased from the prior investigation in favour of higher yield, while the target thicknesses under study have also increased in favour of a more practical design.

Table 1: Proton beam spot sizes

| Proton spot size description | $\sigma_x$ (mm) | $\sigma_y$ (mm) | $\sigma_{x'}$ (mrad) | $\sigma_{y'}$ (mrad) |
|---|---|---|---|---|
| Default | 0.055 | 1.1066 | 0.38 | 0.38 |
| Small | 0.15 | 0.15 | 0.6366 | 0.6366 |

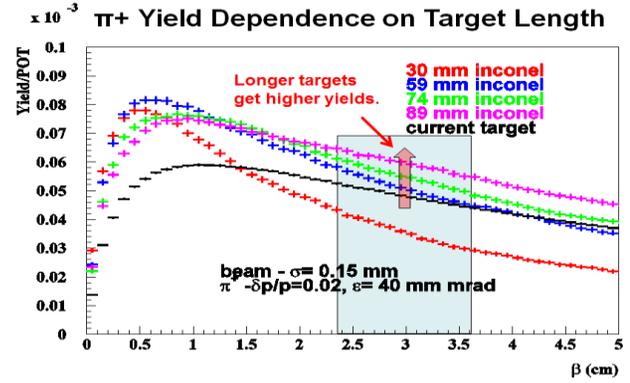

Figure 4: Pion yield in acceptance estimated at the target for targets of different lengths as a function of different values for the β-function [6]. Value for β-function is estimated to be ~2.5 to ~3.5 cm. Radius of each target is 0.375 mm, except for 89 mm long target which has radius of 0.45 mm.

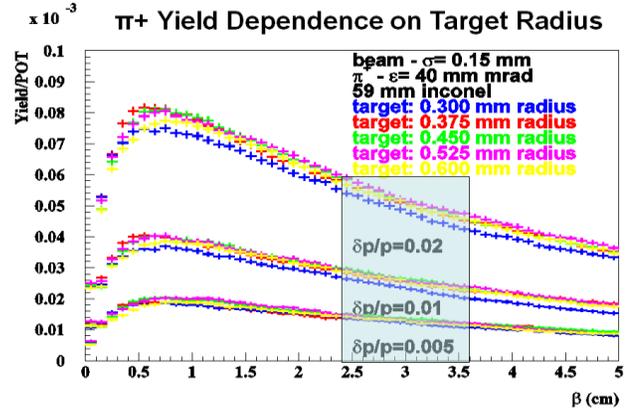

Figure 5: Pion yield in acceptance estimated at the target for targets of different radii as a function of different values for the β-function [6]. Value for β-function is estimated to be ~2.5 to ~3.5 cm.

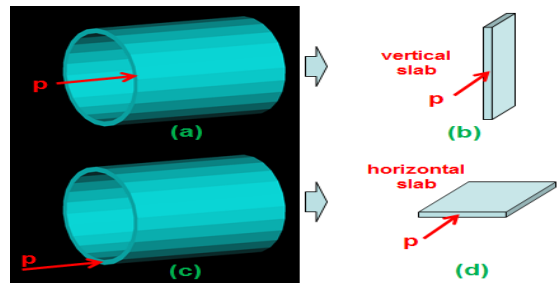

Figure 6: Thin walled cylinder target design to accommodate cooling and the identification between location where proton beam hits the target and the vertical or horizontal slab targets simulated.

Table 2: Dimensions and orientations of targets studied. Material of all targets is Inconel.

| Shape | Length (mm) | Width /Diameter |
|---|---|---|
| Solid Cylinder | 74 | 0.75 |
| Horizontal Slab | 59, 74, 89, 118 | 0.60, 0.75, 1.00, 1.25 |
| Vertical Slab | 59, 74, 89 | 0.60, 0.75, 0.90 |
| Default Pbar Target | chord ~75 | — |

The simulation results are shown in Figure 7, where optimal gains in useful pions can be attributed to two factors:
1. A 61% increase due to the smaller spot size of the proton on the default pbar target.
2. A further 41% enhancement due to a change of target geometry from that of the default pbar target to one of a horizontal slab of length 118 mm and width 0.75 mm. This corresponds to the thin walled cylinder having the wall be 0.75 mm thick and 118 mm long.

The combined yield enhancement is 127% over the default proton spot size on the default pbar target. Note that both increases are made possible by a smaller beam size, since use of a thinner target to minimize absorption of pions requires a proton beam that is narrower than the target ($R_{target} \approx 2.5\sigma_{proton\,beam}$). Hence, there is a risk where movement of the proton beam position may have a greater adverse effect on the pion yield compared to a configuration with a wider beam on a wider target. Also, the lack of target configurations that extend beyond the optimal configuration to indeed verify that it is the optimum highlights the preliminary nature of this study. It is obvious that longer targets should be studied as well as having as many targets in the vertical and horizontal orientations. Our bias was towards the horizontal orientation, since our expectation is to have a higher yield in a configuration where the pions exit the target surface in the non-bend plane. Future studies will remove this bias and test both orientations equally.

The results for the horizontal slab of length 74 mm in **Figure 7** shows an initially unexpected increase in yield for the widest configuration studied, which may possibly be attributed to the widening of the proton beam as it traverses the target to a transverse size that peaks production of pions at the surface in the downstream portion of the target. We expect this effect would be secondary compared to the higher intensity proton beam at the upstream portion of the target that produces pions near the surface of a thinner target. The interaction between the two is complicated further by the change in location of maximum pion production for both phenomena with respect to the focal point as the target dimension is varied. Specifically, requiring the center of the target to be at the focal point necessarily pushes the location of upstream higher rate pion production forward in front of the focal point and is worsened for longer targets, while potentially allowing the secondary mechanism to come into focus. This interplay along with the expectation of low pion production at the end of a long target suggests a change in scheme to extract the optimal target length where the downstream end of the target is fixed to be as close as possible to the Li lens and the front of the target moves upstream as the target is lengthened. Future studies will test this new approach will likely arrive at an optimum with less complications and possibly elucidate if we are indeed seeing the interplay of these two phenomena.

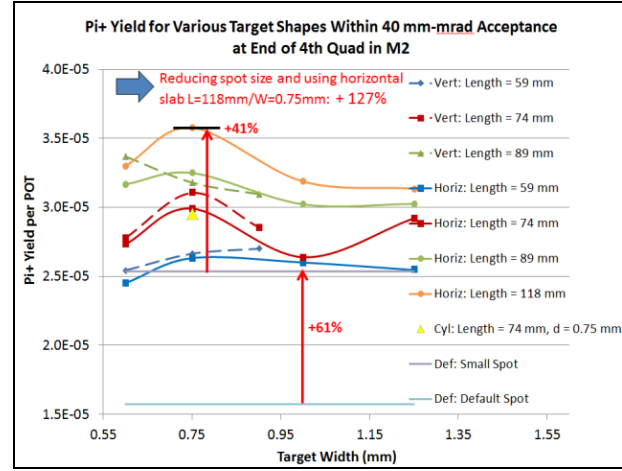

Figure 7: Yield of π+ for targets of various shapes and orientations, plus effect of small spot size of proton on default pbar target.

## ACKNOWLEDGEMENT

The support, encouragement, and contributions of Chris Polly, Mary Convery, and Dean Still of Fermilab are gratefully acknowledged.

## SUMMARY AND FUTURE

A preliminary study of the New g-2 Experiment target system at Fermilab was performed to optimize the yield of useful pions. Reducing the proton beam spot size on the existing pbar target alone increases the pion yield by 61%. An additional 41% enhancement is possible by changing the target into one that has a thin dimension vertically of 0.75 mm and is 118mm long. A thin walled cylinder target satisfies the thin wall constraint as well as provides a means for cooling the target. This work is preliminary in that more configurations need to be simulated to find the optimal one, while also taking into account practical constraints.

## REFERENCES

[1] R. M. Carey et al., http://gm2.fnal.gov/public_docs/proposals/Proposal-APR5-Final.pdf
[2] G. W. Bennett et al., Phys. Rev. D 73 (2006) 072003.
[3] N. Mokhov, g-2 collaboration meeting note (2009).
[4] MARS: N. Mokhov, http://www-ap.fnal.gov/MARS
[5] G4beamline: T. Roberts, http://g4beamline.muonsinc.com
[6] S. Striganov, http://beamdocs.fnal.gov/AD/DocDB/0040/004039/001/g-2-striganov2.pdf